\documentclass[twocolumn,showpacs,superscriptaddress,preprintnumbers,amsmath,amssymb,prc]{revtex4-1}

\usepackage{amsmath}
\usepackage{cases}
\usepackage{amssymb}
\usepackage{CJK}
\usepackage{graphicx}
\usepackage{dcolumn}
\usepackage{bm}
\usepackage{color}
\usepackage{ulem}
\usepackage[pagewise]{lineno}

\begin{document}
\begin{CJK*}{GBK}{}
\title{Investigation of giant dipole resonance in heavy deformed nuclei with the EQMD model}

\author{S. S. Wang}
\affiliation{Shanghai Institute of Applied Physics, Chinese Academy of Sciences, Shanghai 201800, China}
\affiliation{University of Chinese Academy of Sciences, Beijing 100049, China}
\author{Y. G. Ma}\thanks{Email:Corresponding author. ygma@sinap.ac.cn}
\affiliation{Shanghai Institute of Applied Physics, Chinese Academy of Sciences, Shanghai 201800, China}
\affiliation{Shanghai Tech University, Shanghai 200031, China}
\author{X. G. Cao}
\affiliation{Shanghai Institute of Applied Physics, Chinese Academy of Sciences, Shanghai 201800, China}
\author{W. B. He}
\affiliation{Institute of Modern Physics, Fudan University, Shanghai 200433, China}
\author{H. Y. Kong}
\affiliation{Shanghai Institute of Applied Physics, Chinese Academy of Sciences, Shanghai 201800, China}
\affiliation{University of Chinese Academy of Sciences, Beijing 100049, China}
\author{C. W. Ma}
\affiliation{Institute of Particle and Nuclear Physics, Henan Normal
University, Xinxiang 453007, China}

\date{\today}

\begin{abstract}
The deformation evolution of giant dipole resonance (GDR), in the chains of Sm and Nd isotopes, are investigated in the framework of an extended quantum molecular dynamics (EQMD) model. The mass number dependence of resonance peak position $(E_{m})$ in the major and minor axis directions of deformed nuclei  as well as the difference $\Delta E_{m}$ between them are described in detail. The correlation between the splitting ($\Delta E_{m} /\bar{E}_m $) of the GDR spectra and the deformation($ \beta_{2}$) is further studied. The results confirm that $\Delta E_{m} /\bar{E}_m $ is proportional to $ \beta_{2}$. By comparing the calculation with the experimental data on photon absorption cross section $\sigma_{\gamma}$, it shows that the EQMD model can quite well reproduce the shape of GDR spectra from spherical to prolate shape. The GDR shapes in $^{134}$Sm, $^{136}$Sm, $^{138}$Sm, $^{130}$Nd, $^{132}$Nd and $^{134}$Nd are also predicted. In addition, the symmetry energy coefficient $(E_{sym})$ dependence of GDR spectra of $^{150}$Nd is also discussed. It is found that the calculated GDR spectrum in the EQMD model is perfectly consistent with the experimental results when $E_{sym}$ equals to 32 MeV.
\end{abstract}


\maketitle
\end{CJK*}

\section{Introduction}
\label{introduction}
Giant dipole resonance (GDR) is the most prominent characteristic in the excitation spectrum for all nuclei (except for deuterons) in the nuclide chart, which has been regarded as a specific probe for measuring the shape of a nucleus.  Due to this fact, there is increasing interest in applications to the dynamics of exotic nuclei. The relationship of the geometrical and dynamical symmetries of $\alpha$-clustering configuration with the number and centroid energies of peaks in the GDR spectra has been discussed in Ref.~\cite{wbhe2014,wbhe2016}. The evolution of GDR with neutron excess for the neutron-rich oxygen isotopes has been systematically measured in Ref.~\cite{alta2001}. Additionally, since deformation effects in GDR spectrum were firstly seen more than fifty years ago in terms of a double humped photon cross section peak~\cite{egf1962}, it has been well established that the GDR peak is split into two components due to the different frequencies of dipole oscillation along the major axis and minor axis in heavy deformed nuclei~\cite{bsi2011,blhe1975,jmew1995,abbr1998}.

Deformed nucleus provides an interesting testing ground since there is a strong interplay between the structure of the GDR and the ground-state deformation ~\cite{mdan1958}. Many works have been done both theoretically~\cite{msss2016,sgmm2011,kytn2013,kytn2011,jampg2005,bsiv2007,sfeb2012,yew} and experimentally  ~\cite{bsi2011,vmml2006,blhe1975,PCA1974,PCA1971,dpbd2013} to investigate the effects of deformation in GDRs in heavy deformed nuclei. Most of the studies of the GDRs in deformed nuclei have been focused on the dependence of the width at half maximum, peak position and strength of dipole resonance on deformation.

Various microscopic methods have been employed to investigate the GDRs of deformed nuclei such as random-phase approximation approach ~\cite{msss2016,sgmm2011,kytn2013,kytn2011,wkvo2008,jllg2007}, time-dependent Skyrme-Hartree-Fock method~\cite{jampg2005,sfeb2012}, time-dependent density functional theory~\cite{isab2011} and phonon damping model \cite{nddk1999}. The excitation of the GDRs in the experiment is induced by inelastic scattering~\cite{tpug2016,dhyj1976,mihs2002}, photoabsorption ~\cite{vmml2006,PCA1974,PCA1971,vapr2011}, $\gamma$-decay~\cite{jhka1990} and so on. However, few researches have been conducted about the GDRs in heavy deformed nuclei using a dynamical method.

In this article, the EQMD model \cite{tmaru1996} has been applied to study the GDRs in Sm and Nd isotopes. The initial ground-state deformed nuclei are boosted by imposing a dipole excitation to obtain the GDR spectra. Both the brief introduction of EQMD model and the methods of calculating GDR spectrum are shown in Sec.~\ref{modelformalisim}. To check the reliability of our calculation, the evolution of dipole moments in coordinator space and in momentum space versus time are exhibited in Sec.~\ref{resultsanddiscussion}. The discussions, including the mass number dependence of resonance width along the major and minor axis, the comparison of the calculations with the experimental measurement as well as the effect of symmetry energy coefficient on GDR, are also carried out in this section. Finally, Sec.~\ref{summary} gives the summary.
\section{MODEL AND FORMALISM}
\label{modelformalisim}
\subsection{Introduction of the EQMD model}
The extended quantum molecular dynamics (EQMD) model was developed from the quantum molecular dynamics (QMD) model~\cite{qmd,qmd2,qmd3,qmd4}
 by adding the so-called Pauli potential to the effective interaction and treating the width of Gaussian wave packets for each nucleon as a dynamical variable. The initial ground-state nuclei are obtained at their minimum energy states which are sufficiently stable so that they can be considered as at the real ground states ~\cite{tmaru1996,sswc2015,heNT}. Thanks to the advantage, EQMD model has been successfully applied to study the giant dipole resonance (GDR) of the alpha-clustering nuclei~\cite{wbhe2014,wbhe2016}. In this article, we use the EQMD model to investigate the GDRs in heavy deformed nuclei. A brief introduction of the EQMD model is presented as follows.

In the model, the total wave function of the system is treated as a direct product of Gaussian wave packets of all nucleons~\cite{tmaru1996}
\begin{equation}\label{wavefunction}
\Psi =\prod_{i}\varphi \left ( \mathbf{r}_{i} \right ),
\end{equation}
\begin{equation}\label{singlewavefunction}
\varphi\left ( \mathbf{r}_{i} \right ) = \left ({ \frac{ \nu _{i}+
\nu _{i}^{*}}{2\pi}} \right )^{3/4}exp\left[-\frac{\nu _{i}}{2}
\left ( \mathbf{r}_{i}- \mathbf{R}_{i} \right )^{2}+\frac{i}{\hbar}
\mathbf{P}_{i}\cdot \mathbf{r}_{i}  \right],
\end{equation}
where $\textbf{R}_{i}$ and $\textbf{P}_{i}$ are the centers of position and momentum of the $i$th wave packet, respectively. The Gaussian width $\nu_{i}$ is  introduced using a complex as

\begin{equation}\label{wavewidth}
\nu_{i}\equiv\frac{1}{\lambda_{i}}+i\delta_{i},
\end{equation}
where $\lambda_{i}$ and $\delta_{i}$ denote the real and the imaginary parts. They are dynamical variables in the process of initialization.

The effective interaction contains Skyrme, Coulomb, Symmetry potential as well as the Pauli potential.
\begin{equation}\label{effinter}
H_{int}=H_{Skyrme}+H_{Coulomb}+H_{Symmetry}+H_{Pauli}.
\end{equation}

The simplest form is used for the Skyrme interaction
\begin{equation}\label{hsymm}
H_{Skyrme}=\frac{\alpha}{2\rho_{0}}\int\rho^{2}(\textbf{r})d^{3}r+\frac{\beta}{(\gamma+1)\rho_{0}^{\gamma}}\int\rho^{\gamma+1}(\textbf{r})d^{3}r,
\end{equation}
with $\alpha$ =-124.3MeV, $\beta$ =70.5MeV, and $\gamma$ =2. They are obtained by fitting the ground-state properties of the finite nuclei.

For the Pauli potential, a very simple form is applied by introducing a phenomenological repulsive potential which inhibits nucleons of the same spin $S$ and isospin $T$ to come close to each other in the phase space
\begin{equation}\label{Paulipotential}
H_{Pauli}=\frac{c_{p}}{2}\sum _{i}\left ( f_{i}-f_{0} \right )^{\mu }
\theta \left ( f_{i}-f_{0} \right ),
\end{equation}
where $c_{p}$ denotes the strength of the Pauli potential, which equals to 15MeV. For the other two parameters, we take $f_{0}$= 1.0 and $\mu$ =1.3. $f_{i}$ is the overlap of a nucleon $i$ with the same spin $S$ and isospin $T$ as follows
\begin{equation}\label{fi}
f_{i}\equiv\sum _{j}\delta \left ( S_{i}, S_{j} \right )\delta \left ( T_{i}, T_{j} \right )\left |
\left \langle \varphi _{i}\mid \varphi _{j^{}} \right \rangle \right |^{2},
\end{equation}
and $\theta$ is the unit step function.

The symmetry potential is written as
\begin{equation}
H_{Symmetry}=\frac{E_{sym}}{2\rho _{0}}\sum_{i,j\neq i}\int\left[ 2\delta \left ( T_i,T_j \right )-1 \right ]
\rho_i\left ( \mathbf{r} \right )\rho_j\left ( \mathbf{r} \right )d^3r,
\end{equation}
where $E_{sym}$ is the symmetry energy coefficient.

The stability of nuclei in the model description is very important to study the structure effects of nuclei, for example deformation structure. In the EQMD model, the energy-minimum state is considered as the ground state of initial nuclei. At the beginning, a random configuration is given to a nucleus. Then the initial ground-state nucleus are obtained by solving the damped equations of motion as
\begin{equation}\label{initialization}
\begin{split}
\dot{\textbf{R}}_i=\frac{\partial{H}}{\partial{\textbf{P}_i}}+\mu_\textbf{R}
\frac{\partial{H}}{\partial{\textbf{R}_i}}, \qquad & \dot{\textbf{P}}_i=-\frac{\partial{H}}{\partial{\textbf{R}_i}}+\mu_\textbf{P}
\frac{\partial{H}}{\partial{\textbf{P}_i}}, \\
\frac{3\hbar}{4}\dot{\lambda}_i=-\frac{\partial{H}}{\partial{\delta_i}}+
\mu_\lambda\frac{\partial{H}}{\partial{\lambda_i}},\qquad& \frac{3\hbar}{4}\dot{\delta}_i=\frac{\partial{H}}{\partial{\lambda_i}}+
\mu_\delta\frac{\partial{H}}{\partial{\delta_i}}, \\
\end{split}
\end{equation}
where $\mu_\textbf{R}$, $\mu_\textbf{P}$, $\mu_\lambda$, and $\mu_\delta$ are damping coefficients. With negative values of these coefficients, the system goes to its (local) energy minimum point.

\begin{equation}\label{hamlttime}
\begin{split}
\frac{dH}{dt}&=\underset{i}{\sum}[\frac{\partial{H}}{\partial{\textbf{R}_i}}\dot{\textbf{R}}_i
+\frac{\partial{H}}{\partial{\textbf{P}_i}}\dot{\textbf{P}}_i
+\frac{\partial{H}}{\partial{\lambda_i}}\dot{\lambda}_i
+\frac{\partial{H}}{\partial{\delta_i}}\dot{\delta}_i] \\
&=\underset{i}{\sum}[\mu_\textbf{R}(\frac{\partial{H}}{\partial{\textbf{R}_i}})^{2}
+\mu_\textbf{P}(\frac{\partial{H}}{\partial{\textbf{P}_i}})^{2}
+\frac{4\mu_\lambda}{3\hbar}(\frac{\partial{H}}{\partial{\lambda_i}})^2 \\
&+\frac{4\mu_\delta}{3\hbar}(\frac{\partial{H}}{\partial{\delta_i}})^2]\leq 0.
\end{split}
\end{equation}

\subsection{Giant Dipole Resonance in deformed nuclei}

To study the GDRs of the deformed nuclei that have an ellipsoidal shape in the framework of EQMD model, we firstly need to obtain the phase space information of the ground state deformed nuclei, which have been proven and measured in the experiments. Nevertheless the fact is that not all of the phase space distributions of the initial nuclei obtained from initialization with the EQMD model  are ellipsoidal, it is indispensable to select the deformed nuclei from all initial nuclei, whose deformations are consistent with the experimental measurements. Here, the initial deformed nuclei are selected by comparing the calculated deformation parameter $ \beta_{2}$ with the experimental data. $\beta_{2}$ is a parameter linked through the symmetry-axis radius $ R_{x}$ and the radius $ R_{0}$ of the spherical nucleus with the same mass in accordance with the following relationship
\begin{equation}\label{deformpara}
\beta_{2}=\sqrt{\frac{4\pi }{5}}\frac{R_{x}-R_{0}}{R_{0}}.
\end{equation}
Note that in the EQMD model, $R_{0}$ is taken as the root-mean-squared radius.

According to the macroscopic description of GDR given by the Goldhaber-Teller model~\cite{mgo1948}, the GDR is considered as a coherent dipole oscillation of the bulk of protons and neutrons along opposite direction in an excited nucleus. In this work, the initial nucleus is triggered by means of giving a displacement between protons and nucleons at  $t$ = 0 fm/c and then a dipole excitation is triggered and evolved with time. The dipole moments of the system in coordinator space $D_{G}(t)$ and in momentum space $K_{G}(t)$ are defined, respectively, as follows~\cite{wbhe2014,wbhe2016,whl2010,ctyg2013,vbmc2001}:

\begin{equation}\label{coormoments}
D_{G}\left ( t \right )=\frac{NZ}{A}\left [ R_{Z}\left ( t \right )- R_{N}\left ( t \right )\right ],
\end{equation}

\begin{equation}\label{momentunmoments}
 K_{G}\left ( t \right )=\frac{NZ}{A\hbar}\left [ \frac{P_{Z}\left ( t \right )}{Z}-\frac{P_{N}\left ( t \right )}{N} \right ].
\end{equation}
Where $R_{Z}(t)[ P_{Z}(t)]$ and $R_{N}(t)[ P_{N}(t)]$ are the $c.m's$ of the protons and neutrons in coordinate (momentum) space, respectively.

The strength of the dipole resonance of the system at energy $E_{\gamma}$ can be derived from the dipole moments $D_{G}(t)$, i.e.,
\begin{equation}\label{emissionprob}
\frac{dP}{dE_{\gamma }}=\frac{2e^{2}}{3\pi \hbar c^{3}E_{\gamma }}\left |  {D}''\left ( \omega  \right )\right |^{2},
\end{equation}
where $\frac{dP}{dE_{\gamma }}$ can be interpreted as the $ \gamma$ emission probability. $ {D}''\left ( \omega  \right )$ is from the the Fourier transform of the second derivative of $D_{G}(t)$ with respect to time, i.e.,
\begin{equation}\label{Fouriertrans}
{D}''(\omega )=\int_{t_{0}}^{t_{max}}{D_{G}}''(t)e^{i\omega t}dt.
\end{equation}
It needs to be noted that the evolution time can't be infinite long in the realistic calculation and the Fourier transform in the infinite time-range is not reasonable due to the GDR spectrum having the natural width. Moreover, the different final states only affect the width of GDR spectra, which determine the effective width of the Fourier transform in Eq.(~\ref{Fouriertrans}). However, they don't affect the peak position ($E_{m}$) of the resonance maximum. So we take $t_{max}$ = 300fm/c as the final state in this work.

The peak of the GDRs in deformed nuclei splits two parts, while there is only one single peak for the spherical nuclei. In EQMD model, we calculate the GDRs along  $x-$ and $z-$ axis, respectively. Analogy to the superposition of two non-interfering Lorentz lines for statically deformed nuclei (the lower-energy line corresponds to oscillations along the major axis and the higher-energy line along the minor) for fitting the experimental data~\cite{blhe1975}, we take the method of the superposition of two GDR spectra to gain the total resonance strength in a deformed nucleus. The formula is given as
\begin{equation}\label{superposition}
\frac{dP}{dE_{\gamma }}=\sum _{i=1}^{2}\frac{\left ( \frac{dP}{dE_{\gamma }} \right )_{mi}}{1+\frac{\left ( E_{\gamma }^{2}-E_{ mi}^{2} \right )^{2}}{E_{ mi}^{2}\Gamma_{i}^{2}}},
\end{equation}
where $ (\frac{dP}{dE_{\gamma }})_{m} $ is the resonance strength maximum; $ E_{m}$ is the peak position of the resonance maximum; $ \Gamma_{i} $ is the resonance width at half-maximum; and $ i=1,2 $ correspond to the $x-$ and $z-$axis resonance components of the deformed nucleus. Noting that the above three parameters are all obtained by fitting the single GDR spectrum along two axes directions with the Gaussian function, and $x-$axis corresponds to the major axis and $z-$axis is the minor axis of the nucleus in our calculation.

\section{Results and discussion}
\label{resultsanddiscussion}

Nuclei in the region of mass number Z$ = $60 (Nd) and Z$ = $62 (Sm) display a transition from spherical, at the neutron number N close to 82, to prolate ellipsoidal shape. Considering this, the chains of Nd and Sm isotopes are used to study the deformation dependence of GDR spectra in the framework of EQMD model. The experimental data of photon absorption cross sections $ \sigma_{\gamma}$ in the GDR range are extracted from Ref.~\cite{PCA1974} for Sm isotopes and from Ref.~\cite{PCA1971} for Nd isotopes.

\begin{table}[!hbp]
\caption{The deformation parameter $\beta_{2}$, the experimental data are from Ref.~\cite{SRA2001} and the calculation are based on Eq.(\ref{deformpara}).}
\begin{tabular}{p{2.5cm}p{3.2cm}p{2.5cm}}
\hline
\hline
 nuclei & $\beta_{2}$        & $\beta_{2}$       \\
        & exp.~\cite{SRA2001}& calcu.            \\
\hline
 $^{130}$Nd  & 0.37$\pm$0.09      & 0.3586$\pm$0.0109 \\
 $^{132}$Nd  & 0.349$\pm$0.03     & 0.3485$\pm$0.0139 \\
  $^{134}$Nd  & 0.249$\pm$0.025    & 0.2558$\pm$0.007  \\
  $^{142}$Nd   & 0.0917$\pm$0.001   & 0.0941$\pm$0.0136 \\
 $^{144}$Nd   & 0.1237$\pm$0.0006  & 0.1019$\pm$0.0047 \\
  $^{146}$Nd  & 0.1524$\pm$0.0025  & 0.1497$\pm$0.0182 \\
 $^{148}$Nd   & 0.2013$\pm$0.0037  & 0.2133$\pm$0.0035 \\
  $^{150}$Nd   & 0.2853$\pm$0.0021  & 0.2733$\pm$0.0173 \\
 $^{134}$Sm  & 0.366$\pm$0.026    & 0.3576$\pm$0.0053 \\
 $^{136}$Sm  & 0.293$\pm$0.015    & 0.2846$\pm$0.0116 \\
 $^{138}$Sm  & 0.208$\pm$0.017    & 0.2091$\pm$0.0169 \\
 $^{144}$Sm  & 0.0874$\pm$0.001   & 0.0869$\pm$0.0087 \\
 $^{148}$Sm  & 0.1423$\pm$0.003   & 0.1547$\pm$0.0113 \\
 $^{150}$Sm  & 0.1931$\pm$0.0021  & 0.1861$\pm$0.0097 \\
 $^{152}$Sm  & 0.3064$\pm$0.0027  & 0.315$\pm$0.0157  \\
 $^{154}$Sm  & 0.341$\pm$0.002    & 0.3443$\pm$0.0072 \\
\hline
\hline
\end{tabular}
\label{betacomp}
\end{table}

The deformation parameter $\beta_{2}$ as one main parameter of describing the deformed nuclei is treated as a probe to select the ground deformed nuclei from all initial nuclei in this article. In TABLE.~\ref{betacomp}, the $\beta_{2}$ in the chains of Nd and Sm isotopes are shown, including the experimental data from Ref.~\cite{SRA2001} and the calculation based on Eq.(\ref{deformpara}). The statistical errors are  also attached for each nucleus in the table. From the table, we can find the $\beta_{2}$ of our calculations are very close to the experimental values.

\begin{figure}
\includegraphics[width=8.6cm]{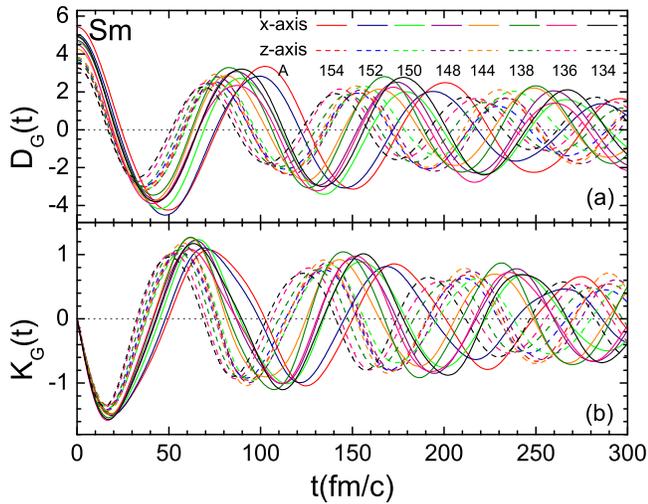}
\caption{(Color online) The time evolution of the dipole moments in coordinator space (a) and in momentum space (b) for the chain of Sm isotopes computed using Eq.~(\ref{coormoments}) and Eq.~(\ref{momentunmoments}) with the EQMD model, respectively. The solid lines denote the $D_{G}$($K_{G}$) along $x-$axis and the dash lines correspond to that along $z-$axis. Dot lines in (a) and (b) (dark gray line) represent $D_G (K_G) = 0$ .}
\label{SmKGDG}
\end{figure}

In this work, the collective motion are divided into two directions along $x-$ and $z-$axis. The initial state wave function of the system is boosted by means of imposing a dipole excitation at $t=0$ fm/c. The time evolution of the dipole moments of Sm isotopes in coordinator space ($D_{G}$) and in momentum space ($K_{G}$) are shown in Fig.\ref{SmKGDG}$(a)$ and $(b)$, respectively. It is found that all the dipole oscillations are symmetrical around  $D_G (K_G) = 0$ except for that close to the initial time. The resonance frequencies along the $x-$axis direction are lower than that in $z-$ axis direction. That is why a deformed nucleus has two splitting peaks in its GDR spectrum. The same situation occurs in the Nd isotopes.

\begin{figure}
\includegraphics[width=8.6cm]{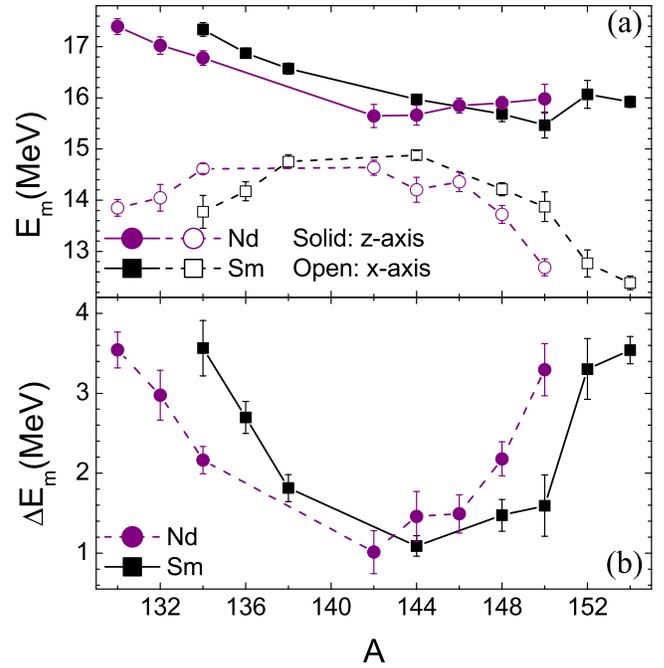}
\caption{(Color online)(a) The peak position ($E_{m}$) of GDRs along $x-$ (open symbols) and $z-$axis (solid symbols) in Sm (black squares) and Nd isotopes (purple circles), respectively. (b) The mass dependence of the difference ($\Delta E_{m}$) between two GDR peak positions in two decomposed directions for each Sm (black squares) and Nd (purple circles) isotope.}
\label{SmNdPk}
\end{figure}

The total GDR spectra are obtained from the superposition of two GDR spectra along $x-$ and $z-$axis by Eq.(\ref{superposition}). The peak position ($E_{m}$) of the resonance maximum and the resonance width ($ \Gamma$) are two indispensable parameters to get the total resonance strength. That is one reason to show the mass evolution of $E_{m}$ of GDRs in the chains of Sm and Nd isotopes in Fig.\ref{SmNdPk}$(a)$. The open symbols denote the $E_{m}$ along $x-$axis, which corresponds to the major axis of a deformed nucleus. And the solid symbols represent the $E_{m}$ in $z-$axis direction, corresponding to the minor axis of the deformed nucleus. It can be seen that with the increasing of $A$, the resonance $E_{m}$ in the $z-$axis direction firstly decreases, and then trends to increase. On the contrary, the resonance $E_{m}$ along $x-$axis direction gradually increases with the increasing of $A$ until the mass number equals to 142 in the chain of Nd isotopes and equals to 144 in the chain of Sm isotopes, and then gradually decreases. It is necessary to note that $^{142}$Nd and $^{144}$Sm are magic nuclei. From Fig.\ref{SmNdPk} $(b)$, it can be easily seen that the distance $\Delta E_{m}$ between the resonance $E_{m}$ along two axis firstly decrease with the increasing of $A$ and then increase, which can also be described as the less the deformation of a nucleus is, the closer the two resonance $E_{m}$ is. It confirms that the splitting peak in the deformed nuclei results from the deformation structure of the nuclei.

\begin{figure}
\includegraphics[width=8.2cm]{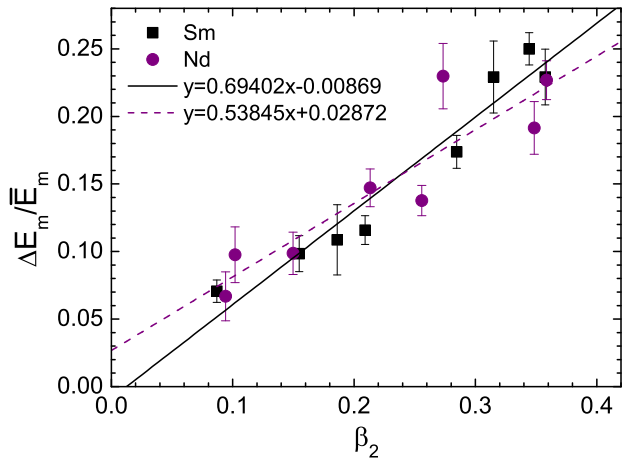}
\caption{(Color online) The correlation between $\Delta E_{m} /\bar{E}_m $ and deformation parameter $\beta_{2}$ which is obtained by Eq.(\ref{deformpara}). Lines are the fitting results. Black squares and purple circles denote the data in the chain of Sm and Nd isotopes, respectively.}
\label{beta2DE}
\end{figure}

Furthermore, the correlation between $\Delta E_{m} /\bar{E}_m $ and deformation parameter $\beta_{2}$ is shown in Fig.\ref{beta2DE}. Note that $\bar{E}_m $ is the mean resonance energy. Black squares denote the data of Sm isotopes and purple circles denote that of Nd isotopes. $\beta_{2}$ is computed based on Eq.(\ref{deformpara}). Both solid and dash lines are the linear fitting results. The fitting parameters also are listed in the figure. For the chain of Sm isotopes, the relationship  between $\Delta E_{m} /\bar{E}_m $ and deformation parameter is $\Delta E_{m} /\bar{E}_m=0.69402 \beta_{2}-0.00869 $; And for the chain of Nd isotopes, the relationship is $\Delta E_{m} /\bar{E}_m=0.53845 \beta_{2}+0.02872 $. It also confirms that the splitting between the $x-$ and $z-$axis modes of the deformed nuclei is proportional to the deformation, which has been described in detail in Ref.~\cite{abbr1998}.

\begin{figure}
\includegraphics[width=8.6cm]{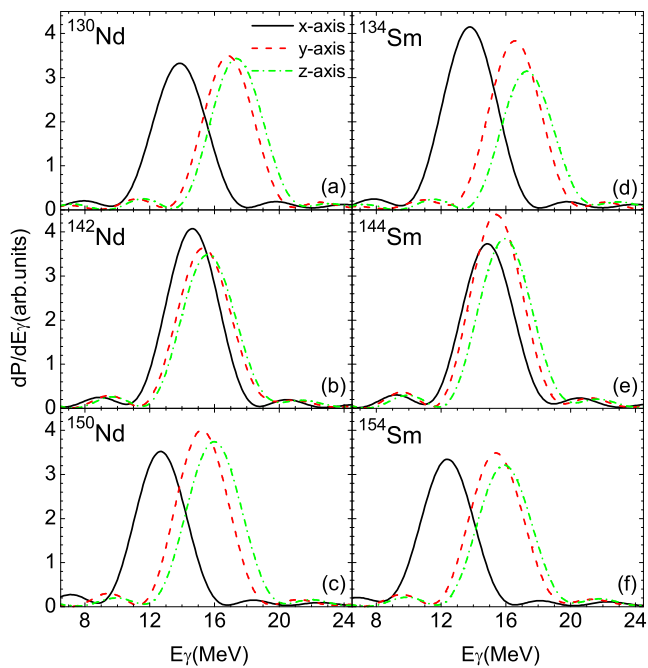}
\caption{(Color online) Dipole strengths in the Nd isotopes and Sm isotopes based on Eq.(\ref{emissionprob}). Solid lines denote mode along $x-$axis direction, dash lines denote mode along $y-$axis direction, and dash dot lines denote mode along $z-$axis direction.}
\label{gdrsingle}
\end{figure}

In Fig.~\ref{gdrsingle}, dipole strengths based on Eq.~(\ref{emissionprob}) for the separate modes in the framework of EQMD model are plotted with different lines. It can be seen that the oscillations along the $x-$axis which denote the major axis of symmetry correspond to the lower-energy state, and that along the $y-$axis and $z-$axis  which denote the minor axis of symmetry correspond to the higher-energy state. For $^{142}$Nd and $^{144}$Sm which are magic nuclei, their resonance peaks along the three axes are so close that the total GDR spectrum have single-hump. However, for $^{130}$Nd, $^{150}$Nd, $^{134}$Sm and $^{154}$Sm, the resonance peaks along the major axis are much smaller than that along the minor axis and the resonance spectra along $y-$axis and $z-$axis nearly overlap. Consequently, in an ellipsoid-deformed nucleus, the total GDR spectrum has two-hump. For the resonance spectra along $y-$axis direction, it is not perfectly identical to the one along the z axis, which mostly results from the shape fluctuation of the initial deformed nucleus in EQMD model.

\begin{figure*}
\includegraphics[width=14.6cm]{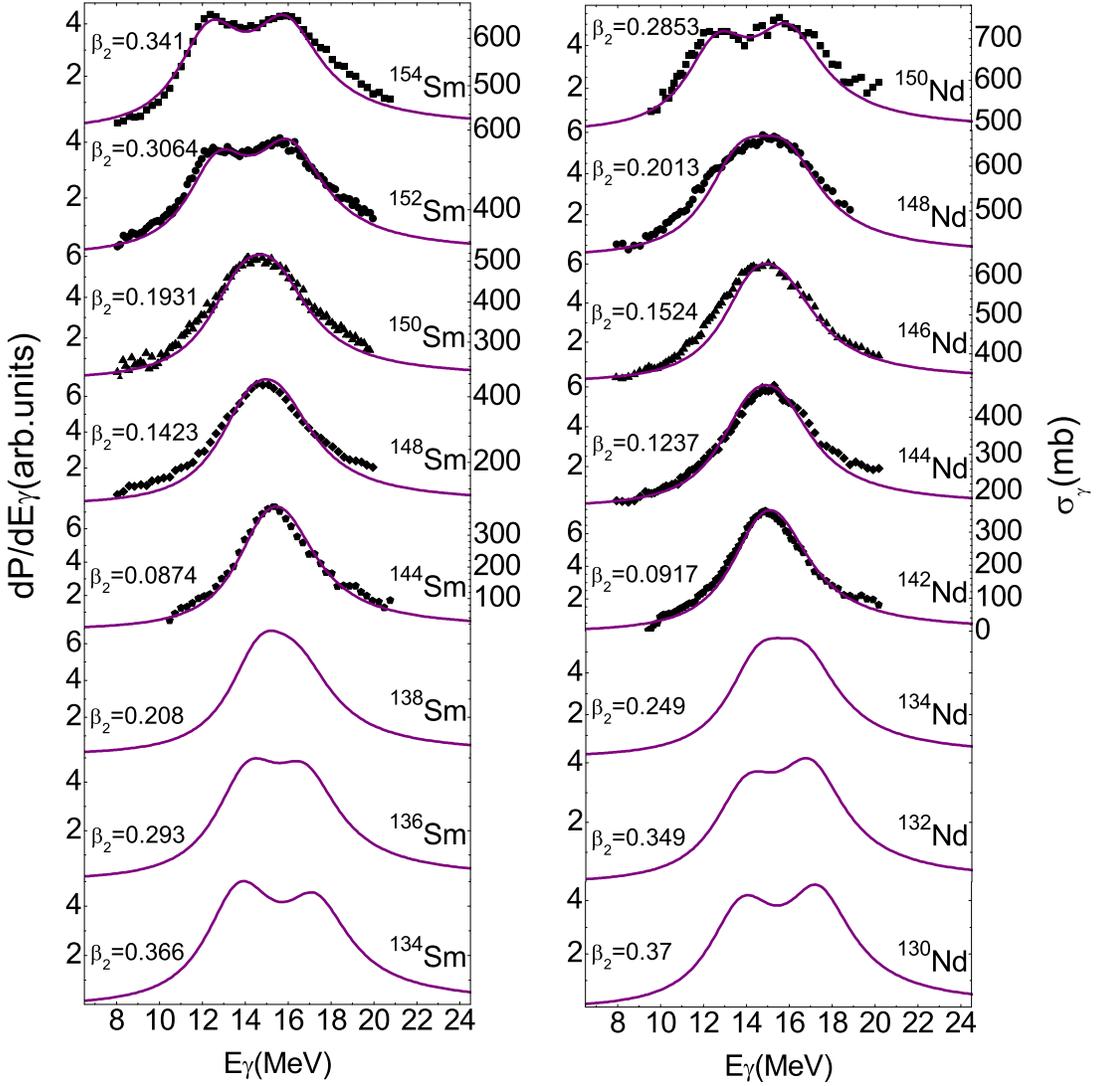}
\caption{(Color online) The GDR spectra in the isotopic chain of Sm ($A = 134-154$) and Nd isotopes ($A = 130-150$). The lines denote the calculation in the EQMD model (scaled by the left $Y$ axis). Dots represent the experimental data (photon absorption cross sections $ \sigma_{\gamma}$, scaled by the right $Y$ axis). The deformation parameter $\beta_{2}$ from Ref.~\cite{SRA2001} are displayed in each panel.}
\label{SmNdgdr}
\end{figure*}

The deformation evolution of the total GDR spectra along $x-$ and $z-$axis directions in Sm and Nd isotopes are plotted in Fig.\ref{SmNdgdr} where the deformation parameter $\beta_{2}$ from Ref.~\cite{SRA2001} are inserted in each panel. The lines denote our calculation by Eq.(\ref{superposition}) in the framework of EQMD model, scaled by the left $Y$ axis and the dots denote the experimental data, scaled by the right $Y$ axis. It needs to be pointed out that the photon absorption cross sections $\sigma_{\gamma}$ in the GDR range for $^{134}$Sm, $^{136}$Sm, $^{138}$Sm, $^{130}$Nd, $^{132}$Nd and $^{134}$Nd are unknown in the experiment so far. From Fig.\ref{SmNdgdr}, one can see that the calculated GDRs can perfectly reproduce the shape of GDR spectra. For example, the GDR spectra have two distinctly splitting peaks in the region of strongly deformed nuclei, such as in $^{154}$Sm, $^{152}$Sm, $^{136}$Sm, $^{134}$Sm, $^{150}$Nd, $^{132}$Nd and $^{130}$Nd, while there is only one maximum value when the deformation of a nucleus is very small, especially for the magic nuclei $^{144}$Sm and $^{142}$Nd. What's more, with the decreasing of the deformation of nuclei, the GDR width also decreases which occurs both in Sm and Nd isotopes. All of these characteristics above have been observed in the experiment. Additionally, it can also be seen that all of the peak position, for the isotopic chain of Sm ($A = 142-154$) and Nd ($A = 140-150$) isotopes, is well consistent with the measured data. Therefore, the results not only confirm the reliability of the methods and the model to study the GDR in deformed nuclei, but also predict the shapes of GDR spectra in Sm ($A = 134,136,138$) and Nd ($A = 130,132,134$) isotopes, which is possible to be verified by experiments.

\begin{figure}
\includegraphics[width=8cm]{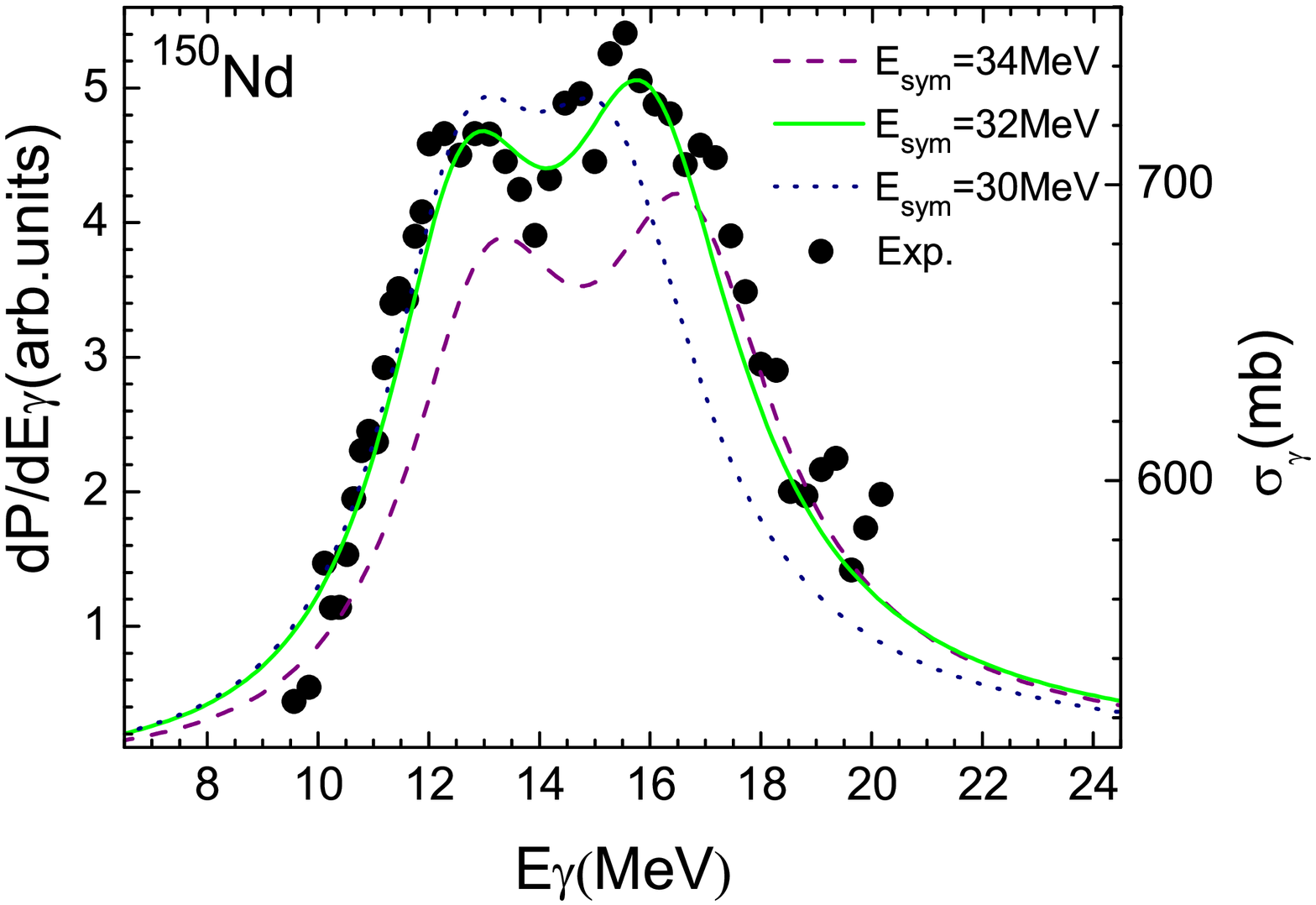}
\caption{(Color online) The dependence of the GDR spectra on symmetry energy coefficient ($E_{sym}$) in heavy deformed nuclei $^{150}Nd$. Dots show the experimental data (photon absorption cross sections $ \sigma_{\gamma}$, scaled by the right $Y$ axis). The dash line, solid line and dot line correspond to the calculating results (scaled by the left $Y$ axis) at $E_{sym}$ equals to 34, 32 and 30 MeV, respectively.}
\label{Nd150Esymdepen}
\end{figure}

The dependence of the GDRs on symmetry energy coefficient ($E_{sym}$) is also discussed for the heavy deformed nucleus of $^{150}$Nd in the EQMD model. The results are shown in Fig.~\ref{Nd150Esymdepen}. The dots represent the measured data from the experiment, which are the photon absorption cross sections $ \sigma_{\gamma}$ in the GDR range, scaled by the right $Y$ axis. The calculations are plotted as the different lines corresponding to different $E_{sym}$, scaled by the left $Y$ axis. From the figure, it is cleanly seen that with the increasing of $E_{sym}$ from 30 MeV to 34 MeV, the GDR spectra of the system have the obvious trend of moving to the right, i.e. the energy position of GDR is governed by the symmetry energy. Meanwhile, one can find that the calculation is well consistent with the experimental at $E_{sym}$ = 32 MeV, which demonstrates that the 32 MeV is the best choice for the symmetry energy coefficient to investigate the GDRs in heavy deformed nuclei in the framework of EQMD model.

\section{Summary and outlook}
\label{summary}

In summary, the deformation evolution of giant dipole resonance, in the isotopic chains of Sm and Nd, has been systematically studied under the framework of an extended quantum molecular dynamics model. The discussions are conducted about the mass dependence of resonance peak position $(E_{m})$ in the major and minor axis directions as well as the difference $\Delta E_{m}$ between them, respectively. The $\Delta E_{m}$ between the two resonance $E_{m}$ firstly decreases with the increasing of $A$ and then trends to increase. It confirms that $\Delta E_{m}$ is extremely sensitive to the deformation of a nucleus. Moreover, the correlation between $\Delta E_{m} /\bar{E}_m $ and the deformation parameter ($ \beta_{2}$) is considered. For the isotopic chain of Sm, $\Delta E_{m} /\bar{E}_m = 0.69402 \beta_{2}-0.00869 $, and for the isotopic chain of Nd, $\Delta E_{m} /\bar{E}_m = 0.53845 \beta_{2} + 0.02872 $. It further confirms that the splitting of the GDR spectra along major axis and minor axis is proportional to the deformation of a nucleus. Additionally, by comparing the calculation with the experimental data of photon absorption cross section $\sigma_{\gamma}$, the results show that EQMD model can perfectly reproduce the shape of GDR spectra. The GDR spectra in $^{134}$Sm, $^{136}$Sm, $^{138}$Sm, $^{130}$Nd, $^{132}$Nd and $^{134}$Nd  are also predicted in detail. Finally, the dependence of GDR spectra of $^{150}$Nd on symmetry energy coefficient $(E_{sym})$ are considered. The results demonstrate that the calculation is well consistent with the experimental results at $E_{sym}$ = 32 MeV. It suggests that the EQMD model can be used to study the configuration structure of deformed nuclei.

In light of the success for describing the deformed GDR by the EQMD model, it is expected that it could also be applied to treat the pygmy dipole resonance (PDR) which can be considered as the oscillation between the weakly bound neutron skin and the isospin neutral proton-neutron core for neutron-rich nuclei. Previously on, a traditional QMD model has shown its capability to investigate PDR and GDR in Ni isotopes by Coulomb excitation \cite{ctyg2013}, it is naturally expected that the EQMD can do an even good job for the PDR study since  more reasonable ground state nuclei could be obtained in the EQMD initialization in contrast to the traditional QMD initialization.

\begin{acknowledgments}

This work is supported by the National Natural Science Foundation of China under Contracts Nos.11421505, 11305239 and 11220101005, the Major State Basic
Research Development Program in China under Contracts No. 2014CB845401, the Strategic Priority Research Program of the Chinese Academy of Sciences under Grant
No. XDB16, the Youth Innovation Promotion Association of CAS No.2017309, the Program for Science and Technology Innovation Talents in Universities of Henan Province under Grant No. 13HASTIT046, Natural Science Foundation
in Henan Province under Grant No. 162300410179.

\end{acknowledgments}

\end{document}